\documentclass[10pt,letterpaper,twocolumn,aps,pra, superscriptaddress,longbibliography]{revtex4-1}

\usepackage{amsmath}    % need for subequations
\usepackage{graphicx}   % need for figures %dblfloatfix
\usepackage{verbatim}   % useful for program listings
\usepackage{color}      % use if color is used in text
\usepackage[usenames,dvipsnames,svgnames,table]{xcolor}
\usepackage{subfigure}  % use for side-by-side figures
\usepackage[hidelinks]{hyperref}   % use for hypertext links, including those to external documents and URLs
\usepackage{multirow}
\usepackage{float}
\usepackage{braket}
\usepackage{siunitx}
\usepackage{bbm}
\usepackage{microtype}       % als letztes -- Verbesserter Textsatz
\usepackage{bibunits}
\usepackage{tabularx}
\usepackage{todonotes}
\graphicspath{{images/}}

\renewcommand{\i}{\mathrm i}
\newcommand{\unit}[1]{\ensuremath{\,\mathrm{#1}}}
\newcommand{\e}[1]{\ensuremath{\mathrm e^{#1}}}

\newcommand{\CZ}{C$Z$-gate}
\newcommand{\CZs}{C$Z$-gates}
\newcommand{\oo}{\ensuremath{\ket{11}}}
\newcommand{\tz}{\ensuremath{\ket{20}}}

\newcommand{\ielem}{\ell}
\newcommand{\iset}{i}
\newcommand{\isetB}{j}
\newcommand{\Iset}{I}
\newcommand{\numset}{n}
\newcommand{\ecmat}{K}
\newcommand{\ecbit}{b}
\newcommand{\ecspin}{z}
\newcommand{\intfactor}{\alpha}
\newcommand{\id}{I}

\renewcommand{\dag}{^{\dagger}}

\newcommand{\puttitle}{Improving the Performance of Deep Quantum Optimization Algorithms\\ with Continuous Gate Sets}

\usepackage{lipsum}

\begin{document}

\title{\puttitle}
\author{Nathan~Lacroix}
\affiliation{Department of Physics, ETH Zurich, CH-8093 Zurich, Switzerland}
\author{Christoph~Hellings}
\affiliation{Department of Physics, ETH Zurich, CH-8093 Zurich, Switzerland}
\author{Christian~Kraglund~Andersen}
\affiliation{Department of Physics, ETH Zurich, CH-8093 Zurich, Switzerland}
\author{Agustin~Di~Paolo}
\affiliation{Institut Quantique and D\'epartement de Physique, Universit\'e de Sherbrooke, Sherbrooke J1K2R1 Qu\'ebec, Canada}
\author{Ants~Remm}
\affiliation{Department of Physics, ETH Zurich, CH-8093 Zurich, Switzerland}
\author{Stefania~Lazar}
\affiliation{Department of Physics, ETH Zurich, CH-8093 Zurich, Switzerland}
\author{Sebastian~Krinner}
\affiliation{Department of Physics, ETH Zurich, CH-8093 Zurich, Switzerland}
\author{Graham~J.~Norris}
\affiliation{Department of Physics, ETH Zurich, CH-8093 Zurich, Switzerland}
\author{Mihai~Gabureac}
\affiliation{Department of Physics, ETH Zurich, CH-8093 Zurich, Switzerland}
\author{Alexandre~Blais}
\affiliation{Institut Quantique and D\'epartement de Physique, Universit\'e de Sherbrooke, Sherbrooke J1K2R1 Qu\'ebec, Canada}
\affiliation{Canadian Institute for Advanced Research, Toronto,  ON, Canada}
\author{Christopher~Eichler}
\affiliation{Department of Physics, ETH Zurich, CH-8093 Zurich, Switzerland}
\author{Andreas~Wallraff}
\affiliation{Department of Physics, ETH Zurich, CH-8093 Zurich, Switzerland}
\date{\today}

%\pacs{}

\begin{abstract}
Variational quantum algorithms are believed to be promising for solving computationally hard problems and are often comprised of repeated layers of quantum gates.
An example thereof is the quantum approximate optimization algorithm (QAOA), an approach to solve combinatorial optimization problems on noisy intermediate-scale quantum (NISQ) systems. 
Gaining computational power from QAOA critically relies on the mitigation of errors during the execution of the algorithm, which for coherence-limited operations is achievable by reducing the gate count. 
Here, we demonstrate an improvement of up to a factor of 3 in algorithmic performance as measured by the success probability, by implementing a continuous hardware-efficient gate set using superconducting quantum circuits. This gate set allows us to perform the phase separation step in QAOA with a single physical gate for each pair of qubits instead of decomposing it into two C$Z$-gates and single-qubit gates.
With this reduced number of physical gates, which scales with the number of layers employed in the algorithm, we experimentally investigate the circuit-depth-dependent performance of QAOA applied to exact-cover problem instances mapped onto three and seven qubits, using up to a total of 399 operations and up to 9 layers. 
Our results demonstrate that the use of continuous gate sets may be a key component in extending the impact of near-term quantum computers.
\end{abstract}

\maketitle

%%%%%%%%%%%%%%%%%%%%%%%%%%%%% MAIN TEXT %%%%%%%%%%%%%%%%%%%%%%%%%%%%%%%%%%%
\section{Introduction}
Quantum computers have the potential to outperform classical computers on a range of computational problems such as prime factoring~\cite{Shor1994} and quantum chemistry~\cite{Cao2019}. Although many of these applications will require quantum error correction~\cite{Lidar2013} to provide a quantum advantage, there is an increasing interest in exploring quantum applications on noisy intermediate-scale quantum (NISQ) devices~\cite{Preskill2018} available in the near-term.
Recent experiments have demonstrated a computational advantage of quantum computers~\cite{Arute2019}, explored many-body physics~\cite{Zhang2017k, Bernien2017} and simulated small-scale quantum chemistry problems~\cite{OMalley2016, Kandala2017, Arute2020}. Moreover, there is a significant interest in solving optimization problems on quantum computers, in particular with the quantum approximate optimization algorithm (QAOA)~\cite{Farhi2014, Otterbach2017, Arute2020a}. 
This variational algorithm has been used to study a range of discrete~\cite{Zhou2018c, Farhi2014, Hadfield2019, Farhi2019} and continuous~\cite{Verdon2019} optimization problems, and may have applications for unstructured search~\cite{Jiang2017a}.
While there is currently no proof that it can provide an asymptotic quantum advantage, QAOA is an emerging approach for benchmarking quantum devices and is a candidate for demonstrating a practical quantum speed-up on near-term NISQ devices. 

To find an approximate solution to a combinatorial problem with QAOA, a problem Hamiltonian is formulated, whose ground state corresponds to the solution of the combinatorial problem. To approximate this ground state, a quantum computer prepares an ansatz state with a parameterized gate sequence, whose parameters are iteratively updated by a classical optimizer.
The gate sequence consists of layers, each characterized by two variational parameters, $\gamma_q$ and $\beta_q$, see Fig.~\ref{fig1}. The number of layers, $p$, sets the depth of the algorithm and QAOA can reach the global optimum of any cost function for $p\to\infty$~\cite{Farhi2014}. It is therefore expected that the computational power of QAOA increases with $p$. 
In practice, however, the number of layers that can be executed reliably on near-term quantum computers is limited due to finite gate errors induced by relaxation, dephasing and pulse imperfections~\cite{Alam2019, Zhou2018c}. 

Small-scale implementations of QAOA, while restricted to solving problems that can also be efficiently solved on classical computers, provide crucial insights into the feasibility and challenges related to the execution of the algorithm on NISQ devices. 
Previous studies of QAOA with superconducting qubits~\cite{Otterbach2017, Matsumine2019, Alam2019, Bengtsson2019, Arute2020a}, photonics~\cite{Qiang2018} and trapped ions~\cite{Pagano2019} highlight the applicability of QAOA on a range of platforms and illustrate the breadth of problems that can be addressed with QAOA.
The work presented in Ref.~\cite{Otterbach2017} studied the MaxCut problem, which is the canonical problem for QAOA~\cite{Farhi2014}, with up to 19 qubits, Ref.~\cite{Matsumine2019} studied a channel decoding problem, Ref.~\cite{Pagano2019} searched the eigenstate of all-to-all connected Ising models with up to 40 qubits and Ref.~\cite{Bengtsson2019} considered an exact-cover problem with 2 qubits. 
Many of these experiments consider problems that can be solved with shallow QAOA circuits ($p=1$ or $2$). However, these examples may not be representative of the broad range of problems that can be addressed with QAOA. Indeed, studies of all-to-all connected Ising models show that deep circuits may be needed~\cite{Arute2020a}. 

When implementing quantum algorithms on a quantum device, it is common to decompose the gate sequence into a discrete set of gates available on the hardware.
To improve performance, recent experiments have explored continuous gate sets motivated by applications in quantum simulations~\cite{Barends2015, Roushan2016}, quantum chemistry~\cite{Ganzhorn2019, Foxen2020} and for QAOA using $XY$ interactions~\cite{Abrams2019a}.
In this work, we benchmark QAOA with a continuous hardware-efficient gate set.
We present a controlled arbitrary-phase gate (C-ARB gate), which allows to execute each QAOA layer with only one two-qubit gate per $ZZ$-term in problem Hamiltonians formulated as Ising models, see Fig.~\ref{fig1}(a). We demonstrate how our gate set shortens the QAOA sequence and, thus, leads to better performance for a fixed QAOA depth compared to a decomposed implementation of the algorithm with a discrete gate set. 
In particular, we demonstrate with two concrete examples that the reduction in gate sequence duration outweighs errors originating from the interpolation of parameters necessary for implementing the continuous gate set.
Taking advantage of this gain in performance, we investigate the trade-off between experimental noise, which favors shallow circuits, and increasing the number of layers, which is needed to solve complex problem instances. 
 
\section{Implementation}
\label{sec:impl}
The objective function of many NP-complete discrete optimization problems can be mapped to an Ising Hamiltonian~\cite{Karp1972, Lucas2014},
\begin{equation}
\hat{C} = \sum_{i<j} J_{ij} Z_{i}Z_{j} + \sum_{i=1}^n h_i Z_{i},
\label{eq:HamC}
\end{equation}
where $Z_i$ is the Pauli-$Z$ operator for spin $i$.
QAOA can find the ground state of this Hamiltonian by minimizing the expectation value of $\hat{C}$ for the ansatz state $\ket{\vec\gamma, \vec\beta}$ where $\vec\gamma = (\gamma_1, \ldots, \gamma_p)$,  $\vec\beta = (\beta_1, \ldots,\beta_p)$ are variational parameters. 
In particular, the quantum circuit preparing $\ket{\vec\gamma, \vec\beta}$ consists of $p$ layers each containing a phase-separation operator $U_{C}=\e{-\i \gamma_q \hat{C}}$ and a mixing operator $U_B=\e{-\i \beta_q \hat B}$, where $\hat B = \sum_{i} X_{i}$, with $q=1,\ldots,p$~\cite{Farhi2014}. Since all terms of $\hat{C}$ commute, we can implement each term $U_C^{ij} = \e{-\i \frac{\Gamma_{ij}}{2}Z_i Z_j}$ separately, where $\Gamma_{ij}/2 = \gamma_q J_{ij}$ is a continuous parameter.

\begin{figure}[t] % !b H
\centering
\includegraphics[width=\linewidth, ]{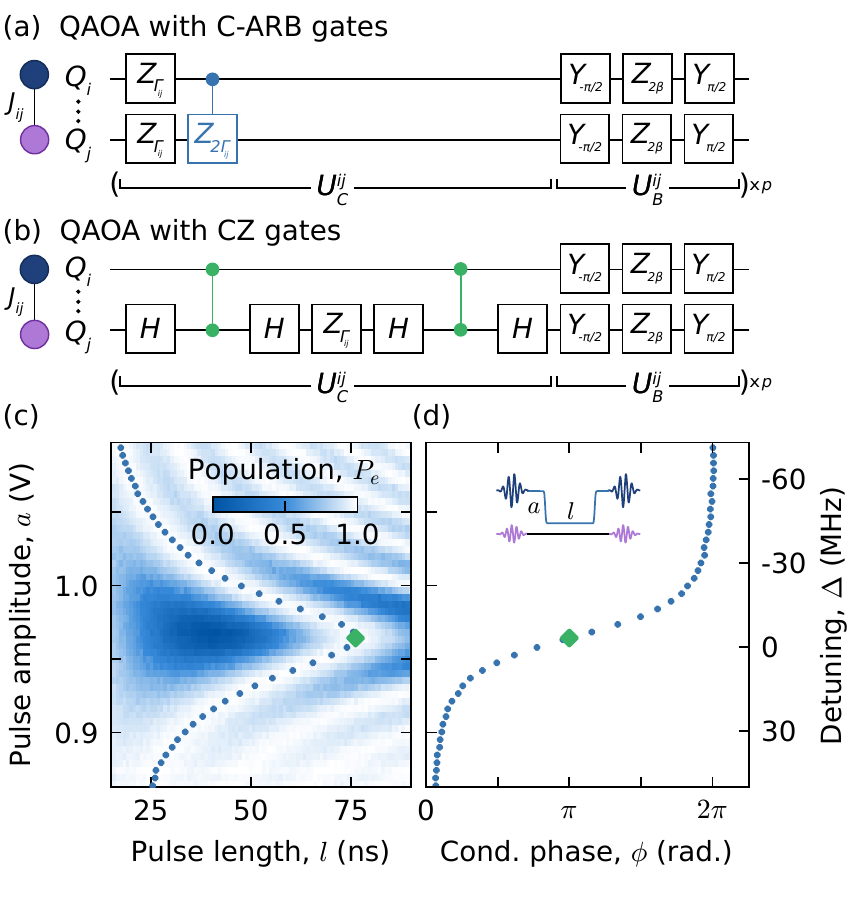}
\caption{(a)~Quantum circuit of a layer $q$ of QAOA for the two-qubit subspace $\ket{Q_iQ_j}$, using the controlled arbitrary-phase gate (blue) to rotate the \oo{} state by an angle $2\Gamma_{ij}$ where $\Gamma_{ij} = 2\gamma_qJ_{ij}$. (b)~A QAOA layer with the phase-separation unitary $U^{ij}_C$ decomposed into C$Z$ gates (green) and additional Hadamard gates and single-qubit $Z$-gates. (c)~Excited-state population $P_e$ of the control qubit $Q_i = A_1$ brought in interaction with the target qubit $Q_j = B_2$ via a flux pulse. We perform a two-dimensional sweep of flux pulse amplitude $a$ and flux pulse length $l$, and indicate the maximum population recovery with blue dots. (d) Conditional phase for the dots indicated in (c). The green diamond corresponds to the \CZ. The right axis indicates the detuning between \oo{} and \tz{}. The inset depicts the pulse sequence used to measure the conditional phase. Single-qubit $\pi$-pulses and $\pi/2$-pulses are shown in dark blue and purple, respectively. The flux pulse (light blue) of amplitude $a$ and length $l$ is applied to the control qubit $Q_i$, see Appendix~\ref{app:arb} for more details.}
\label{fig1}
\end{figure}

A common approach is to decompose $U_C^{ij}$ into a gate sequence consisting of two conditional phase rotations of $\pi$, i.e.\ standard {\CZs}, combined with several single-qubit gates~\cite{Otterbach2017, Bengtsson2019}.
We present such a decomposition in Fig.~\ref{fig1}(b), where the dependence on the continuous parameters $\Gamma_{ij}$ is introduced via an arbitrary-angle single-qubit $Z$-rotation. An alternative approach is to use a single controlled arbitrary-phase gate (C-ARB gate), which can add any desired phase factor $\e{-\i \phi}$ to the \oo{} state. This gate naturally applies the angle $2\Gamma_{ij}$ and, together with two single-qubit $Z$-rotations, realizes the unitary $U_C^{ij}$, see Fig.~\ref{fig1}(a).

In QAOA, the number of unitaries $U_C^{ij}$ grows linearly with the number of two-qubit terms in $\hat C$ and with the number of QAOA layers, $p$. Thus, it is essential that each $U_C^{ij}$ is implemented with high fidelity. The direct implementation we present in this work significantly reduces both the physical gate count and the sequence duration. Thus, this approach is expected to find correct solutions to complex problems with higher probability. 

\begin{figure}[t] % !b H
\centering
\includegraphics[width=\linewidth]{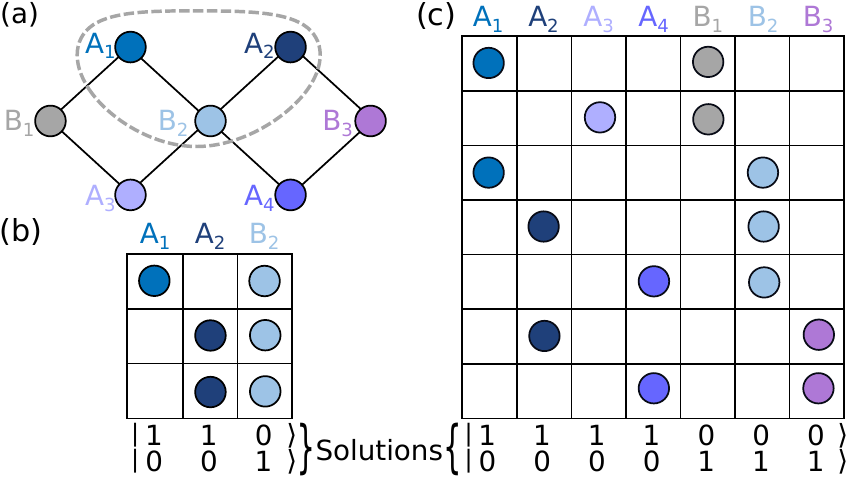}
\caption{(a)~Hardware connectivity graph of the quantum device. Dots correspond to qubits and edges indicate between which pairs of qubits two-qubit gates can be realized. The grey dashed line indicates the subset of qubits used for the three-qubit problem instance depicted in (b). (b)~Visual representation of the incidence matrix $\ecmat$ (dots indicating entries $\ecmat_{\ielem\iset}=1$) for a chosen three-qubit exact-cover problem instance. The labels above the columns (and the colors) indicate which physical qubits are used to represent the corresponding subset. The two solution states are indicated below the grid. (c)~Visual representation of the incidence matrix for a chosen seven-qubit problem instance. }
\label{fig:problem}
\end{figure}

We run QAOA on a quantum device with 7 superconducting transmon qubits, see Appendix~\ref{app:device} for device parameters and a false-colored micrograph of the device.
The qubits are pairwise connected as illustrated in Fig.~\ref{fig:problem}(a). Single-qubit $X$ and $Y$-rotations are implemented with microwave pulses, while $Z$-rotations are performed as virtual gates~\cite{McKay2016a} which take zero time as they are implemented through a redefinition of the reference frame.
To realize \CZs, we use a standard approach relying on a flux pulse which shifts the transition frequency of one of the qubits to bring the $\ket{11}$ state of a pair of coupled qubits in resonance with the non-computational $\ket{20}$ state~\cite{Strauch2003, DiCarlo2009, Barends2014}.
The resulting hybridization leads to a coherent population oscillation between the two states.
The frequency detuning between the \oo{} and \tz{} states, $\Delta$, is 0 during the gate, and after an interaction of the duration corresponding to one oscillation period, the population returns to the \oo{} state with an added phase of $\pi$, see green diamond in Fig.~\ref{fig1}(c) and (d). 
We generalize the \CZ{} to a C-ARB gate on our device by exploiting near-resonant interactions of the \oo{} and \tz{} states~\cite{Barends2015}, i.e.\ $\Delta\neq 0$, to acquire conditional phase angles ranging from 0 to $2\pi$, see Fig.~\ref{fig1}(d).
We vary $\Delta$ by sweeping the flux pulse amplitude and simultaneously adapting the pulse length to maximize population recovery in the computational subspace, see blue dots in Fig.~\ref{fig1}(c). Details about the gate implementation are provided in Appendix~\ref{app:arb}.

We compare the performance of both approaches on two example instances of the NP-complete exact-cover problem~\cite{Karp1972}.
The aim of exact cover is to decide whether it is possible to cover all elements in a set $S$ \emph{exactly once} by an appropriate selection of subsets $\lbrace V_\iset\rbrace$ from a given collection of subsets $V$.
In the example visualized in Fig.~\ref{fig:problem}(b), each row corresponds to an element of a three-element set $S$, while each column corresponds to a subset $V_\iset$ out of three given subsets.
The dots visualize which elements (rows) are included in a subset (column). In this picture, the task is to find a selection of columns such that each row is covered by exactly one dot.
This condition is fulfilled by two solutions: selecting the first two columns or selecting the last column. 
In a mathematical formulation of the exact-cover problem (see Appendix~\ref{app:cover}), the grid in Fig.~\ref{fig:problem}(b) corresponds to a visual representation of an incidence matrix $\ecmat$,
where a dot in row $\ielem$ and column $\iset$ indicates an entry $\ecmat_{\ielem\iset}=1$ while empty cells in the grid indicate entries equal to $0$.
When mapping an instance of exact cover to an Ising Hamiltonian~\cite{Lucas2014, Vikstal2019}, the $\iset$-th qubit encodes whether a subset $V_\iset$ is selected or not, see Appendix~\ref{app:cover}.
In the visualization in Fig.~\ref{fig:problem}(b), the qubit that represents a subset is indicated by the label above the column and by the color used for the dots.
Fig.~\ref{fig:problem}(c) shows an example of a larger instance of exact cover with seven subsets, requiring seven qubits.

To focus on the comparison between the two methods for realizing the two-qubit unitaries $U_C^{ij}$, these two problem instances are chosen such that the resulting Ising Hamiltonians respect the hardware connectivity graph of our device, see Fig.~\ref{fig:problem}(a), and that all single-qubit terms vanish, i.e. $h_i = 0$. 
The three-qubit problem instance depicted in Fig.~\ref{fig:problem}(b) yields an Ising Hamiltonian with $J_{A_1B_2} = 0.5$ and $J_{A_2B_2} = 1$. 
In the basis $\ket{A_1A_2B_2}$, the two possible selections of columns covering all rows, namely $\mathcal{A} = \lbrace A_1, A_2\rbrace$ and $\mathcal{B} =\lbrace B_2 \rbrace$, are encoded with the states $\ket{1 1 0}$ and $\ket{0 0 1}$, respectively, where a 1 in position $\iset$ indicates that the $\iset$-th column of $\ecmat$ is included in the selection of subsets. 
For the seven-qubit problem instance of Fig.~\ref{fig:problem}(c), we have $J_{A_3B_2} = 0$ and $J_{ij} = 0.5$ for all other physically connected qubit pairs. This instance also possesses two solutions, $\mathcal{A}= \lbrace A_1, A_2, A_3, A_4\rbrace$ and $\mathcal{B} = \lbrace B_1, B_2, B_3 \rbrace$, corresponding to the states $ \ket{1111000}$ and $\ket{0000111}$, respectively, using the basis $\ket{A_1A_2A_3A_4B_1B_2B_3}$.
Note that we have labeled the qubits in Fig.~\ref{fig:problem}(a) such that the solutions always correspond to either selecting the qubits labeled with $A$ or the qubits labeled with $B$, see Appendix~\ref{app:cover}.

The QAOA circuit solving the three-qubit problem instance consists of 15 (32) operations per layer while the corresponding circuit for the seven-qubit problem instance consists of 42 (98) operations per layer, for the direct (decomposed) implementation, respectively. 
Thus, the seven-qubit problem instance, which requires deep QAOA circuits, yields a circuit comprising 259 (399) operations in total for the direct (decomposed) implementation at $p=6$ ($p=4$), see Appendix~\ref{app:pulse} for more details.

\section{Performance of QAOA}
A single-layer QAOA implementation ($p=1$) is a useful intermediate benchmark towards implementing multi-layer QAOA circuits since there are only two variational parameters $\vec\gamma = (\gamma_1)$ and $\vec\beta=(\beta_1)$, hereafter referred to as $\gamma$ and $\beta$ for ease of notation, which allows us to map out the full optimization landscape experimentally.
As further discussed in Appendix~\ref{app:symmetries}, when $p=1$, the cost-function landscape is $\pi/2$-periodic in $\beta$ for a problem without single-qubit terms. Moreover, since all eigenvalues of $\hat C$ are odd multiples of $\frac{1}{2}$, the landscape is $2\pi$-periodic in $\gamma$. Finally, the landscape is always point-symmetric around the center point of a period. We can thus reduce our considerations to $\gamma\in[0,\pi[$ and $\beta\in[0,\pi/2[$.  
For each pair of parameters, we prepare the state $\ket{\gamma, \beta}$ 20000 times, see Appendix~\ref{app:pulse} for the full pulse sequence, and we evaluate the cost function $C(\gamma,\beta) = \bra{\gamma, \beta} \hat C \ket{\gamma, \beta}$. 
We use a three-level readout scheme discussed in Appendix~\ref{app:device}, which allows us to discard the measurement outcomes with leakage outside of the computational space, see Appendix~\ref{app:post_selection}. 
In the context of QAOA, discarding leakage events corresponds to reducing the effective number of shots available for evaluating the cost function by rejecting outcomes that are not valid bit-strings.
In this regard, leakage is different from other undetectable errors, for which such a post-selection cannot be done.

We observe that the resulting cost-function landscapes, see Fig.~\ref{fig:landscape}, are odd functions of $\beta$ with a line symmetry axis at $\beta = \pi/4$, see Appendix~\ref{app:symmetries}.
The locations of all extrema in the measured landscape, see Fig.~\ref{fig:landscape}(a), are in good agreement with noise-free simulations, see Fig.~\ref{fig:landscape}(b), which suggests that the coherent errors are small in our implementation. 
Errors due to decoherence mostly affect the contrast of the landscapes~\cite{Xue2019a}, see Fig.~\ref{fig:landscape}(c) and Appendix~\ref{app:sims}. The distortions of the local extrema located at $\gamma > \pi/2$ are attributed to the residual $ZZ$-coupling between the qubits~\cite{Krinner2020}, which we confirm with master-equation simulations, see Appendix~\ref{app:sims}. 

\begin{figure}[t] % !b H
\centering
\includegraphics[width=\linewidth]{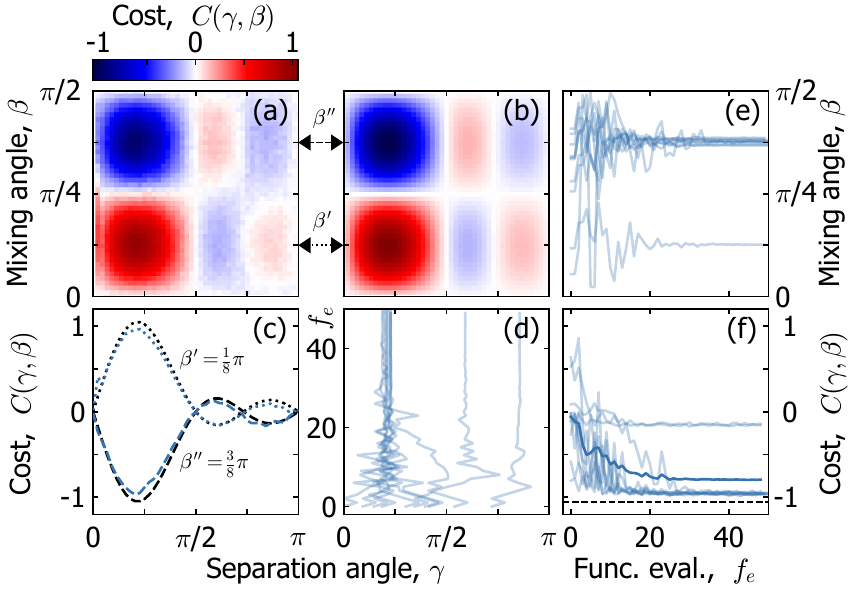}
\caption{Cost function evaluated for $p=1$ on the three-qubit problem instance, using C-ARB gates. (a) Cost-function landscape as a function of variational parameters measured with direct implementation of C-ARB gates. (b) Cost-function landscape obtained from noise-free simulations. (c) Experimental evaluation (blue) and simulation (black) of the cost function for two horizontal line cuts of (a) and (b), with $\beta'=\pi/8$ (dotted lines) and $\beta''=3\pi/8$ (dashed lines) respectively. (d,e) 10 convergence traces of the separation angle and the mixing angle, respectively, for end-to-end optimization starting from random parameter initialization. (f) Average energy (solid blue line) and individual convergence traces (faded lines) of the energy corresponding to parameters shown in (d,e). }
\label{fig:landscape}
\end{figure}

By embedding the evaluation of $C(\gamma,\beta)$ measured on the quantum device into a classical Nealder-Mead optimizer, we demonstrate that the landscape is suitable as cost function for a classical optimizer.
The closed-loop classical optimizer finds the optimal parameters for most random initialization parameters, see Fig.~\ref{fig:landscape}(d-f), however, some convergence traces get trapped in local minima.
Note that in this single-layer implementation, the cost never reaches the ground-state energy $C_{\textrm{gs}} = -1.5$, neither in the measurement nor in the noise-free simulation, which indicates that QAOA circuits of larger depth are indeed required for this problem.

\begin{figure}[b] % !b H
\centering
\includegraphics[width=\linewidth]{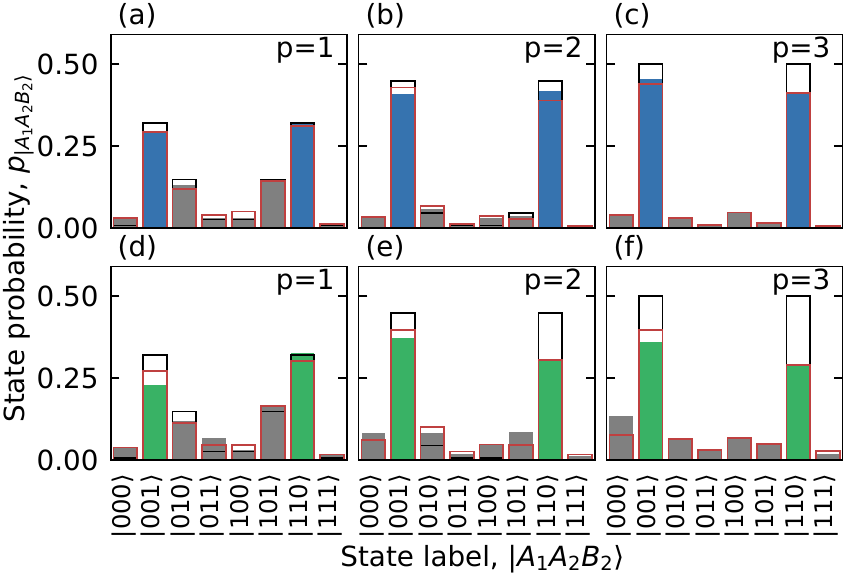}
\caption{Output state probability distribution for the three-qubit problem instance implemented with controlled arbitrary phase gates (a,b,c), and decomposed using \CZs{} (d,e,f). States are measured at optimal parameters for depth of $p=1$ (a,d), $p=2$ (b,e), and $p=3$ (c,f). The filled bars correspond to the measured state probabilities in which we highlight the problem solutions in blue (direct implementation) and green (decomposed implementation), respectively. The black wire-frames are the expected QAOA outcome from noise-free simulations and the red wire-frames are from master-equation simulations.}
\label{fig:state_prob}
\end{figure}

To obtain better approximate solutions to the combinatorial problem instance, we execute QAOA circuits with additional layers and study the effect of the depth $p$ on the output state distribution. To investigate the performance of the quantum part of QAOA rather than the performance of the classical optimizer, we initialize the algorithm with optimal parameters obtained from noise-free simulations. 
We then optimize these parameters locally to correct for small coherent errors, and we estimate the resulting state distribution as a function of depth from 20000 single-shot measurements, see Fig.~\ref{fig:state_prob} for the three-qubit case.
Three layers are required in noise-free simulations (black wire-frames) to fully concentrate the probability distribution on the two solution states $\ket{110}$ and $\ket{001}$ corresponding to the selection of subsets $\mathcal{A}$ and $\mathcal{B}$, respectively.

We quantify the experimental outcomes using the classical fidelity~\cite{Bhattacharyya1943} between the output state probability distribution arising from the measurements, $P$, and from noise-free simulations, $\tilde{P}$,
\begin{equation}
    \mathcal{F}(P,\tilde{P}) = \left(\sum_i \sqrt{P_i} \sqrt{\tilde{P}_i}\right)^2
\end{equation}
where $P_i$ and $\tilde{P}_i$ correspond to the probabilities of the $i$-th basis state in the Hilbert space. Note that $0 \leq \mathcal{F}(P,\tilde{P}) \leq 1 $ with $\mathcal{F}(P,\tilde{P}) = 1$ if and only if $P = \tilde{P}$. 
The state distributions of the implementation using C-ARB gates, see filled bars in Fig.~\ref{fig:state_prob}(a-c), have fidelities of $98.93\unit{\%}$, $95.93\unit{\%}$, and $86.20\unit{\%}$ for $p=1$, 2, and 3 respectively, with respect to the corresponding distribution obtained with noise-free simulations. 
As expected, the reduction of the fidelity with the number of layers $p$ illustrates the accumulation of errors in circuits of increasing depth. However, the concentration of probability on solution states as $p$ increases is stronger than the detrimental effect of the additional errors, such that overall, the probability of measuring a solution increases with $p$. 
By contrast, in the implementation using \CZs, see Fig.~\ref{fig:state_prob}(d-f), the concentration of probability on solution states only compensates the additional errors for $p=2$ while the errors outweigh the gain of an additional layer for $p=3$. This is also reflected by lower fidelities of $96.37\unit{\%}$, $87.43\unit{\%}$, and $64.52\unit{\%}$ for $p=1$, 2, and 3 respectively, and is explained by the fact that decoherence and residual $ZZ$-coupling accumulate over the longer gate sequence.

Master-equation simulations (red wire-frames) are in excellent agreement with the measured distributions, see Appendix~\ref{app:sims} for details. We confirm from these simulations that decoherence is the main limitation in this experiment while residual $ZZ$-coupling cause additional errors, in particular for the decomposed implementation. 

The landscape and state probability distributions of the seven-qubit problem instance presented in Appendix~\ref{app:7qb_landscapes_and_probs} lead to a similar conclusion, i.e.\ the direct implementation exploiting C-ARB gates is in better agreement with noise-free simulations than the decomposed implementation.

\begin{figure}[b] % !b H
\centering
\includegraphics[width=\linewidth]{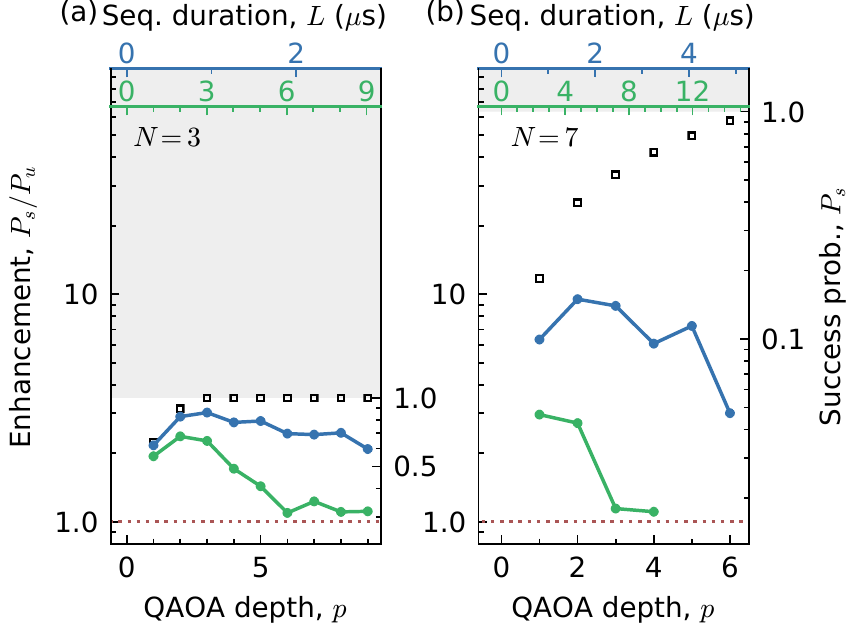}
\caption{Performance of QAOA for (a) three qubits and (b) seven qubits. The blue points are implemented with the direct controlled arbitrary-phase gate and the green points are implemented with the gate sequences decomposed into $CZ$ gates. The black squares indicate the highest success probabilities found with noise-free simulations. The top axes indicate the sequence duration for the direct implementation and the decomposed implementation in green and blue, respectively. The gray areas indicate success probabilities above 1.}
\label{fig:success}
\end{figure}

It is expected that additional layers in QAOA circuits increase the number of reachable states, thereby leading to better approximate solutions in the absence of experimental noise. To gain further insight into the trade-off between the extended reachable state space and the additional noise resulting from increased depth, we determine the enhancement of the success probability provided by the output state distribution over a uniform state distribution as a function of $p$ for both problem instances, see Fig.~\ref{fig:success}.
We define the enhancement as $P_s/P_u$, where $P_s$ is the success probability, i.e.\ the sum of the probabilities of all solution states, and $P_u = 2/(2^N-1)$ is the probability of sampling a solution from a uniform probability distribution over all possible states. Note that we exclude the state $\ket{0}^{\otimes N}$, which is never a solution in the context of exact cover.
We indicate the sequence duration for both implementations with additional axes in Fig.~\ref{fig:success}, where sequence duration is defined as the time between the start of the initialization pulse and the start of the readout. 

In the three-qubit case, Fig.~\ref{fig:success}(a), the direct (decomposed) implementation shows a maximal enhancement of success probability of 3 (2.4) at $p=3$ ($p=2$). The direct implementation (blue dots) provides a higher enhancement than the decomposed implementation (green dots) because the sequence duration is shorter for a fixed $p$, and the problem instance requires at least $p=3$ to reach maximal enhancement in an ideal setting (black squares). 

When the problem increases in complexity, the number of layers required to reach maximal enhancement in a noise-free scenario also increases. For the seven-qubit instance, we find that $p=6$ is required to reach a success probability above 90\%, see Fig.~\ref{fig:success}(b). Consequently, the ability to execute more layers in shorter time provides an even more pronounced advantage. In particular, for the direct implementation, we find that increasing $p$ from 1 to 2 increases the enhancement of the success probability to 9.5. However, when further increasing to $p=3$, the extended reachable state space does not compensate the additional noise arising from the increased sequence duration. For the decomposed version, going beyond $p=1$ does not provide any benefits. 
Thus, for the seven-qubit problem we only benefit from adding layers when taking advantage of the directly implemented C-ARB gates, which improves the performance by a factor of 3 compared to the decomposed implementation.

Finally, to emphasize that the limitations for deeper circuits are directly related to the increased sequence duration rather than the depth itself, we notice that for a fixed sequence duration of $L=5\,\mu$s, both implementations of the seven-qubit instance show similar enhancement of success probability despite being of depth $p=6$ (direct) and $p=2$ (decomposed).

\section{Discussion}
In this work, we show that controlled arbitrary phase gates (C-ARB gates) enable a significant reduction of the number of physical gates required to implement QAOA circuits of any depth on quantum hardware. We demonstrate the advantage of this approach by comparing it to a standard QAOA decomposition on two problem instances of the exact-cover problem, with three and seven qubits, respectively. Despite a more demanding calibration scheme requiring interpolation of gate parameters, C-ARB gates in QAOA circuits systematically outperform the decomposed alternative for a fixed depth and are able to benefit from the extended reachable state space of more layers. 

We foresee an even more pronounced advantage for larger-scale combinatorial optimization problems because the number of layers required to solve problems with QAOA is expected to scale with the number of qubits involved in the experiment~\cite{Bravyi2019, Farhi2020}, in particular for dense problem graphs. In addition, the number of physical two-qubit gates saved within each layer also scales with the number of two-qubit terms in the cost Hamiltonian.
Our results demonstrate that hardware-efficient gate sets are key components in extending the impact of near-term quantum applications, which may become even more relevant when solving problem instances that do not match the connectivity of the hardware. For example, it has recently been observed that the need for swap-gates can significantly reduce the performance of a QAOA implementation if a decomposed implementation of swap gates is used~\cite{Arute2020a}. A direct, hardware-efficient implementation combining a controlled arbitrary phase and a swap-gate may therefore be another key component to improve the performance in these cases, and should be considered in future research.

\section*{Acknowledgments}
The authors are grateful for valuable feedback on the manuscript by B.R.~Johnson and D.~Schuster and for valuable discussions with A.~Choquette-Poitevin and P.~Vikst\aa l.
The authors acknowledge J.~Heinsoo for early contributions towards the realization of C-ARB gates and S. Storz, F. Swiadek and T. Zellweger for their contributions to the measurement setup.

The authors acknowledge financial support by the EU Flagship on Quantum Technology H2020-FETFLAG-2018-03 project 820363 OpenSuperQ, by the Office of the Director of National Intelligence (ODNI), Intelligence Advanced Research Projects Activity (IARPA), via the U.S. Army Research Office grant W911NF-16-1-0071, by the National Centre of Competence in Research Quantum Science and Technology (NCCR QSIT), a research instrument of the Swiss National Science Foundation (SNSF), by the SNFS R'equip grant 206021-170731 and by ETH Zurich. This work was undertaken thanks in part to funding from NSERC, Canada First Research Excellence Fund and ARO W911NF-18-1-0411. The views and conclusions contained herein are those of the authors and should not be interpreted as necessarily representing the official policies or endorsements, either expressed or implied, of the ODNI, IARPA, or the U.S. Government.

%%%%%%%%%%%%%%%%%%%%%% SUPPLEMENTARY MATERIAL %%%%%%%%%%%%%%%%%%%%%%%%%%
\begin{appendix}

\begin{figure}[b] % !b H
\centering
\includegraphics[width=\linewidth]{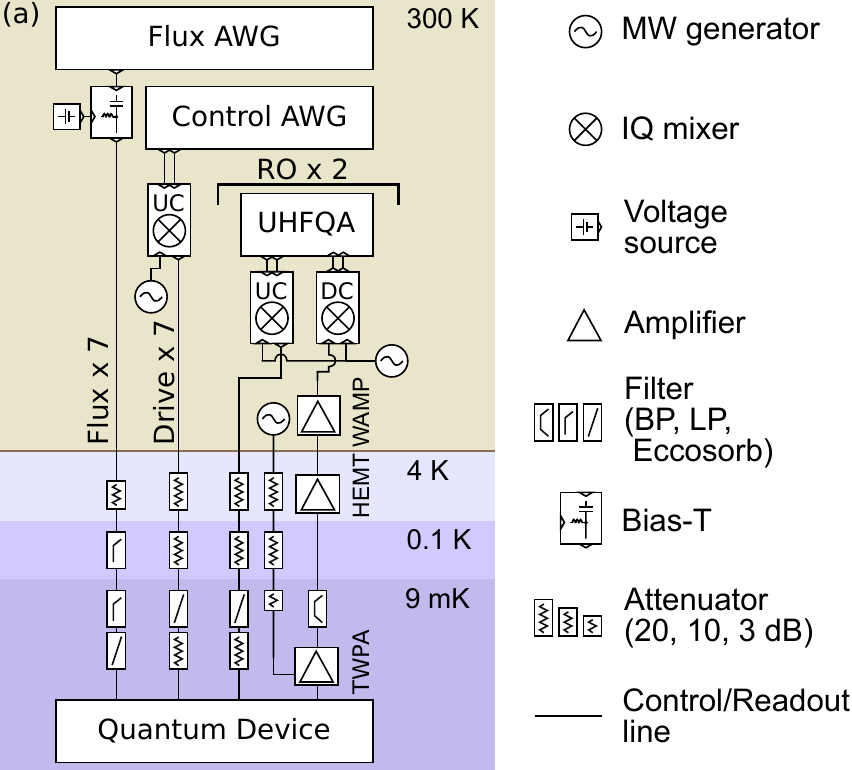}
\caption{Experimental setup. The experiment is controlled by AWGs whose signals are routed to the quantum device through a series of bandpass filters (BP), lowpass filters (LP) and Eccosorb filters. The flux pulses are combined with a voltage source using a bias-T. The IQ signal from the control AWG is upconverted (UC) to a microwave signal using an IQ mixer. The readout signal is generated by the UHFQA, and the output from the quantum device is amplified by a chain of amplifiers before being downconverted (DC) and analyzed by the UHFQA.}
\label{fig:setup}
\end{figure}

\begin{figure}[t] % !b H
\centering
\includegraphics[width=\linewidth]{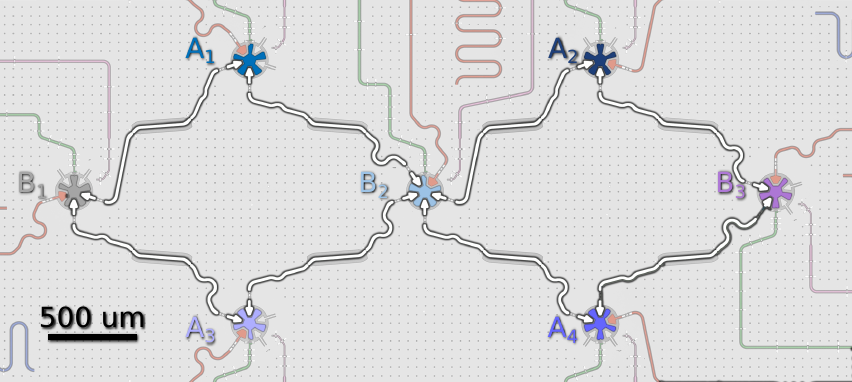}
\caption{Optical micrograph of the device. Each qubit is colored corresponding to the Fig.~\ref{fig:problem} and connected to neighboring qubits by coupling resonators (white). Each qubit is connected to a readout resonator (red), a flux line (green) and a drive line (pink).}
\label{fig:chip}
\end{figure}

\section{Experimental Setup and Device Parameters}
\label{app:device}

\begin{table*}[t]
\centering
\caption{Measured parameters of the seven qubits.}
\begin{tabular}{lccccccc}
\hline
\noalign{\vskip 1mm}
 & A1 & A2 & A3  & A4 & B1 & B2 & B3\\
 \hline
 \hline
Qubit frequency, $\omega_q/2\pi$ (GHz) & 5.462 & 5.684 & 4.077 & 4.195 & 4.825 & 4.920 & 5.165 \\
Lifetime, $T_1$ ($\mu$s) & 12.9 & 6.7 & 24.5 & 21.1 & 16.7 & 14.8 & 14.3 \\
Ramsey decay time, $T_2^*$ ($\mu$s) & 18.1 & 12.2 & 7.4 & 4.6 & 27.2 & 24.5 & 13.3 \\
Readout frequency, $\omega_r/2\pi$ (GHz) & 6.611 & 6.836 & 5.832 & 6.063 & 6.255 & 6.042 & 6.300\\
Readout linewidth, $\kappa_{\mathrm{eff}}/2\pi$ (MHz) & 7.5 & 10.6 & 6.0 & 7.2 & 17.3 & 10.9 & 11.0 \\
Dispersive shift, $\chi/2\pi$ (MHz) & -2.5 & -2.5 & -0.75 & -1.0 & -1.25 & -2.4 & -2.0 \\
Thermal population, $P_{\mathrm{th}}$ ($\%$) & 0.04 & 0.01 & 0.2 & 0.8 & 0.3 & 0.04 & 0.2 \\
$\ket{0}$ readout assignment prob. (\%) & 99.98 & 99.97 & 96.54 & 96.85 & 99.47 & 99.92 & 99.97 \\
$\ket{1}$ readout assignment prob. (\%) & 98.06 & 96.19 & 90.39 & 88.94 & 97.70 & 94.45 & 98.20 \\
$\ket{2}$ readout assignment prob. (\%) & 96.63 & 89.68 & 78.90 & 80.86 & 95.18 & 94.89 & 96.96 \\
\hline
\end{tabular}
\label{tab:qb_params}
\end{table*}

The experiments described in this manuscript are performed in a cryogenic setup~\cite{Krinner2019, Andersen2019b}, the wiring scheme of which is summarized in Fig.~\ref{fig:setup}. Each qubit is controlled by a flux line for frequency tuning which enables two-qubit gates, and a microwave drive line for realizing single-qubit gates. The pulses are generated with arbitrary waveform generators (AWGs). The drive pulses are generated at an intermediate frequency of 100~MHz and upconverted to microwave frequencies. Multiplexed readout is performed via two feedlines~\cite{Heinsoo2018, Andersen2019b} with the readout pulses generated by an ultra-high frequency quantum analyzer (UHFQA). The measurement signals at the output ports of the sample are first amplified with a wide-bandwidth near-quantum-limited traveling-wave parametric amplifier (TWPA)~\cite{Macklin2015}, then with a high-electron-mobility transistor (HEMT) amplifiers and finally with low-noise, room-temperature amplifiers (WAMP). Thereafter, the signals are downconverted and processed using the weighted integration units of the UHFQAs.

The quantum device~\cite{Andersen2019b} shown in Fig.~\ref{fig:chip}, is fabricated on a high-resistivity intrinsic silicon substrate. Photolithography and reactive ion etching are used to define resonators, signal lines and qubit structures in a 150$\,$nm thin niobium film sputtered onto the substrate. We also add air bridges to the device to establish a well-defined ground plane and for cross-overs in signal lines. The Al/AlOx/Al Josephson junctions of the transmon qubits are fabricated using electron-beam lithography and shadow evaporation.

The parameters of the device listed in Table~\ref{tab:qb_params}, are measured using standard spectroscopy and time-domain methods. For the readout, we characterize the ability to identify the correct qubit state as well as the second excited state of each transmon qubit. In particular, we use two weighted-integration units per qubit to distinguish $\ket{0}$ from $\ket{1}$ and $\ket{1}$ from $\ket{2}$, respectively. A standard Gaussian mixture model is then used to classify the resulting integrated weights of each single shot measurement. 

\begin{figure}[t] % !b H
\centering
\includegraphics[width=0.48\textwidth]{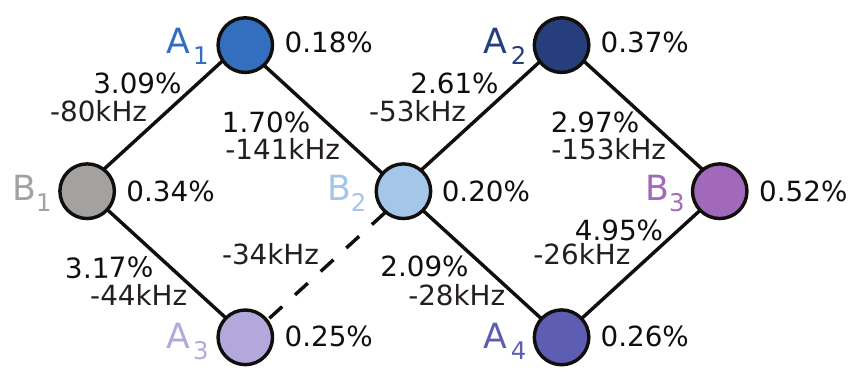}
\caption{Single-qubit error per gate (vertices) and two-qubit error per gate (edges) in percent, measured with randomized benchmarking. We indicate the residual $ZZ$-coupling between each pair of coupled qubits below the corresponding two-qubit gate infidelity. The gate between $A_3$ and $B_2$ is not needed for the problem instances considered in this work (dashed line).}
\label{fig:rb}
\end{figure}

To characterize the gate performance, we perform randomized benchmarking on all qubits to find the error per single-qubit Clifford, see Fig.~\ref{fig:rb}. For the two-qubit gates, we only characterize with a fixed conditional phase of $\pi$ such that the two-qubit gate is in the Clifford group and we measure the error per gate from interleaved randomized benchmarking, see infidelities next to lines indicating the coupling elements in Fig.~\ref{fig:rb}.

We also characterize the residual $ZZ$-coupling, $\alpha_{ij}$, between pairs of coupled qubits, as the frequency shift of qubit $i$ when qubit $j$ is in the excited state. We verify that $\alpha_{ij} = \alpha_{ji}$. The experimentally determined residual $ZZ$-couplings are shown below each error rate in Fig.~\ref{fig:rb}.  

\section{Controlled Arbitrary Phase Gate}
\label{app:arb}
The goal of a C-ARB gate is to apply a unitary in a two-qubit subspace which adds a desired phase $\phi$ to the \oo{} state, 
\begin{equation} \label{eq:carb_unitary}
    U(\phi)=
    \begin{pmatrix}
{1} & {0} & {0} & {0} \\
{0} & {1} & {0} & {0} \\
{0} & {0} & {1} & {0} \\
{0} & {0} & {0} & {\e{-\i \phi}}
\end{pmatrix}
\end{equation}
We realize this unitary by exploiting near-resonant interactions of the \oo{} and \tz{} states. In particular, to span a conditional phase in the range $[0, 2\pi[$, we vary the frequency detuning $\Delta$ between the \oo{} and \tz{} states with a flux pulse of amplitude $a$ and length $l$. During the flux pulse, the population of the $\ket{11}$ state is transferred to the $\ket{20}$ state and after a time $\tau_g$, the population returns to the $\ket{11}$ state with a phase
\begin{equation}
    \phi = \pi \left( 1 + \frac{\Delta}{\sqrt{4J^2 + \Delta^2}}\right),
\end{equation}
where $J$ is the constant coupling strength between the \oo{} and \tz{} states.

To calibrate the flux pulse amplitudes and lengths, we start by measuring the population in the \oo{} state as a function of pulse amplitude and length, yielding a characteristic Chevron pattern (Fig.~\ref{fig1}(c)). For each measured amplitude, we fit the \oo{} population to a cosine to extract the pulse length maximizing the population recovery in the computational subspace. 
Next, we measure the conditional phase for 45 flux pulse amplitudes. For each amplitude, the control qubit is brought to the excited state with a $\pi$-pulse while the target qubit is brought to a superposition state with a $\pi/2$-pulse. Then, a flux pulse is applied to the control qubit and finally a $\pi$-pulse and a $\pi/2$-pulse are applied to the control and the target qubit, respectively. We extract the conditional phase  by varying the phase of the second $\pi/2$-pulse and comparing the phase of the target qubit with and without the initial $\pi$-pulse applied to the control qubit. 
We use the calibrated pulse lengths to ensure high population recovery in the computational subspace for each amplitude. We interpolate the pulse length linearly in between calibration points if required to reach all phases in the range 0 to $2\pi$. Note that on our device, we acquire phase from $0$ to $-2\pi$ but reverse the sign in Fig.~\ref{fig1}(d) for convenience. 
Finally, the flux-biased qubit also acquires a dynamic phase $\phi_D$~\cite{DiCarlo2009}. We compensate for this single-qubit phase shift with a virtual Z-gate after the C-ARB gate. To calibrate $\phi_D$, we compare the phase of the qubit with and without the flux pulse, for amplitudes in the range of interest. 
The number of calibration points is gate-dependent and is set such that we can unwrap the dynamic phase as a function of flux pulse amplitude without ambiguity. This allows us to interpolate (with cubic splines) the dynamic phase between calibration points. 
Gates between specific pairs of qubits also include additional flux pulses on neighboring qubits to avoid undesired interactions, see Appendix~\ref{app:pulse}. We calibrated the dynamic phase acquired by these neighboring qubits simultaneously to the dynamic phase of the qubit directly involved in the gate. 

The calibration procedures are automated such that human interaction is only required to verify the quality of the fits. The approach is thus scalable to larger devices. Note that gate architectures allowing the acquisition of conditional phase as a linear function of the flux pulse length could further simplify and speed up the calibration procedure~\cite{Collodo2020}.

\section{Exact Cover to Ising}
\label{app:cover}
The exact-cover problem is mathematically formulated as follows \cite{Karp1972}.
Given a collection of subsets $V=\{V_\iset\}_{\iset\in{1,\dots,\numset}}$ with $V_\iset\subseteq S$, the task is to verify whether there exists a set of indices $\Iset\subseteq\{1,\dots,\numset\}$ such that
$\{V_\iset\}_{\iset\in \Iset}$ forms a partition of $S$, i.e., the sets in $\{V_\iset\}_{\iset\in \Iset}$ are disjoint and their union equals $S$.
This is the case if
\begin{equation}
0 = \min_{(\ecbit_1,\dots,\ecbit_\numset)\in\{0,1\}^{\numset}} ~ \sum_\ielem \left( 1 - \sum_{\iset} \ecmat_{\ielem\iset}\ecbit_\iset \right)^2
\end{equation}
where the element $\ecmat_{\ielem\iset}$ of the incidence matrix $\ecmat$ is $1$ if the $\ielem$-th element of $S$ is contained in subset $V_\iset$ and $0$ otherwise.
A bit value $\ecbit_\iset=1$ indicates that the subset $V_\iset$ is selected.
Using spins $\ecspin_\iset\in\{\pm1\}$ instead of bits $\ecbit_\iset = \frac{\ecspin_\iset+1}{2}$, multiplying out, and dropping additive constants, the optimization problem can be formulated as \cite{Choi2010a,Lucas2014,Vikstal2019}
\begin{equation}
\min_{(\ecspin_1,\dots,\ecspin_\numset)\in\{\pm1\}^{\numset}} ~ \sum_{\iset<\isetB} J_{\iset\isetB} \ecspin_\isetB \ecspin_\iset + \sum_{\iset} h_{\iset}  \ecspin_\iset 
\end{equation}
where
\begin{align}
\label{eq:J}
J_{\iset\isetB} &= \sum_\ielem \frac{\ecmat_{\ielem\iset}\ecmat_{\ielem\isetB}}{2} \\
\label{eq:h}
h_{\iset} &= \sum_\ielem \ecmat_{\ielem\iset} \left(-1  + \frac{1}{2}\sum_{\isetB} \ecmat_{\ielem\isetB}  \right) .
\end{align}
Solving this optimization problem is equivalent to finding the ground-state energy of the Ising Hamiltonian, see Eq.~\eqref{eq:HamC},
with $J_{\iset\isetB}$ and $h_{\iset}$ values given by Eq.~\eqref{eq:J} and Eq.~\eqref{eq:h}, respectively.
In the visual representations of the incidence matrices depicted in Fig.~\ref{fig:problem}(b) and (c), the bullets represent the entries with $\ecmat_{\ielem\iset}=1$ while empty cells correspond to $\ecmat_{\ielem\iset}=0$. By substituting these values of $\ecmat_{\ielem\iset}$ into the above equations, we see that $h_{\iset} = 0$ for all $\iset$ in both problem instances, and we obtain the values of $J_{\iset\isetB}$ given in Section~\ref{sec:impl}.

To run QAOA without requiring swaps, we need all-to-all physical connectivity between qubits that occur jointly in any row of the incidence matrix $\ecmat$.
For the physical connectivity graph shown in Fig.~\ref{fig:problem}(a), this means that each row can contain only up to two nonzero entries.
The positive sign in Eq.~\eqref{eq:J} reveals that the spins corresponding to a row with two nonzero entries have an antiferromagnetic coupling. This is in line with the exact-cover constraint, which requires that exactly one of them is selected in a valid solution, but not both.
Thus, any problem instance that does not require swaps on our device and that does not decompose into a set of isolated subgraphs
must correspond to a lattice of antiferromagnetically coupled spins.
Then, either all qubits labeled with $A$ or all qubits labeled with $B$ have to be in an excited state in a valid solution.
In the presence of a row (or rows) with a single nonzero entry, some external field term(s) $h_{\iset}$ of the Ising Hamiltonian become(s) nonzero and the solution that fulfills the exact-cover condition also for this row (these rows) is favored.
Otherwise, both solutions are valid, which is the case for the problem instances considered in our experiments.

\section{Properties of QAOA Landscapes}
\label{app:symmetries}
Following Ref.~\cite{Farhi2014}, the parameter $\gamma_q$ can be restricted to $[0, 2\pi[$ if the problem Hamiltonian $\hat{C}$ has integer eigenvalues,
while $\beta_q$ can always be restricted to $[0, \pi[$.
In this appendix, we discuss further periodicity and symmetry properties of the QAOA cost function,
which enable us to reduce the parameter space and better understand the cost-function landscapes we observe.
To this end, we consider the cost of a $p$-layer QAOA circuit,
\begin{equation}
\label{eq:symmetries:base}
C(\vec\gamma',\vec\beta') = \bra{+} U(\vec\gamma',\vec\beta')^\dagger\, \hat C \, U(\vec\gamma',\vec\beta')\ket{+}
\end{equation}
in which $U(\vec\gamma',\vec\beta') = \prod_q \e{-\i \beta_q' \hat{B}} \e{-\i \gamma_q' \hat{C}}$ is the $p$-layer QAOA unitary with $\vec\gamma'=(\gamma_1',\dots,\gamma_p')$ and $\vec\beta'=(\beta_1',\dots,\beta_p')$. 

If the eigenvalues of $\hat{C}$ are integer multiples of $\intfactor$, then by setting $\gamma_q' = \gamma_q+\frac{2\pi}{\intfactor}$ in Eq.~\eqref{eq:symmetries:base} and noting that $\e{\pm\i \frac{2\pi}{\intfactor} \hat{C}}=\id$ is the identity, we find that $C(\vec\gamma',\vec\beta')$ is $(2\pi/\intfactor)$-periodic in $\gamma_q$.

In addition, if all eigenvalues of $\hat{C}$ are odd multiples of $\intfactor$, we have $\e{\pm\i \frac{\pi}{\intfactor} \hat{C}}=-\id$,
where the minus sign is a global phase, so that the cost is $(\pi/\intfactor)$-periodic.
For both problem instances considered in this work, the eigenvalues of $\hat{C}$ are odd multiples of $1/2$, so that the landscapes are $2\pi$-periodic.

Inserting $\beta_q' = \beta_q + \pi$ into Eq.~\eqref{eq:symmetries:base} and noting that $\e{\pm\i \pi \hat{B}}=\prod_\iset \e{\pm\i \pi X_\iset}=\prod_\iset(-\id)$ yields the $\pi$-periodicity in $\beta_q$ mentioned in \cite{Farhi2014}.
Moreover, since $\e{-\i \frac{\pi}{2} \hat{B}}=\prod_\iset \e{-\i \frac{\pi}{2} X_\iset}$ corresponds to an $X_\pi$ rotation of all qubits,
setting $\beta_p' = \beta_p + \frac{\pi}{2}$ in the last layer $p$
corresponds to flipping the sign of all spins before estimating the energy of $\hat C$.
If the Ising Hamiltonian $\hat C$ does not contain single-qubit terms ($h_\iset=0$ for all $i$),
this sign flip does not change the energy, and the cost landscape is $\frac{\pi}{2}$-periodic in $\beta_p$.
As this applies to the examples considered in this paper, we measure the landscapes for $p=1$ only up to $\beta=\beta_1=\frac{\pi}{2}$.

By simultaneously setting $\gamma_q' =  -\gamma_q$ and $\beta_q'=-\beta_q$ in Eq.~\eqref{eq:symmetries:base}, and noting that $\hat C$, $\hat B$, and $\ket{+}$ are real-valued,
we have $C(-\vec\gamma', -\vec\beta')=(C(\vec\gamma, \vec\beta))^\dagger=C(\vec\gamma, \vec\beta)$.
Therefore, the cost landscape is point-symmetric with respect to the origin, which implies that it is also point-symmetric with respect to the center point of a period.
When measuring a landscape, we can thus restrict either $\beta$ or $\gamma$ to half a period without losing information about the landscape.
In the examples shown in this paper, we restrict $\gamma$ to half a period, i.e.\ to the interval $[0,\pi[$.

Finally, when choosing $\gamma_q' = -\gamma_q$ and $\beta_q'=\beta_q$ in Eq.~\eqref{eq:symmetries:base}, we obtain
\begin{equation}
\label{eq:symmetries:odd}
- C(-\vec\gamma, \vec\beta) =
 \bra{+} U'(\vec\gamma,\vec\beta)^\dagger (-\hat C) U'(\vec\gamma,\vec\beta) \ket{+}
\end{equation}
where $U'(\vec\gamma,\vec\beta) = \prod_q \e{-\i \beta_q \hat{B}} \e{-\i \gamma_q (-\hat{C})}$.
This is equivalent to the QAOA cost function for a problem Hamiltonian $\hat C' = -\hat C$.
Thus, in cases for which running QAOA with $\hat C$ and with $-\hat C$ leads to the same landscape,
the landscape is an odd function of $\vec \gamma$.
In particular, this occurs for both problem instances considered in this work.

Due to the point-symmetry observed above, the landscape is also an odd function of $\vec\beta$ if it is an odd function of $\vec\gamma$.
In the landscape plots for $p=1$, this manifests as line symmetries (with a change of the sign of the energy) about both coordinate axes and with respect to the center line of each period.
Within the chosen range of $\beta$, we observe this type of symmetry with respect to the horizontal line $\beta=\frac{\pi}{4}$.

\section{Master-Equation Simulations}
\label{app:sims}

We model the dynamics of our system by a master-equation given by
\begin{align}
\dot{\rho} = -\frac{\i}{\hbar} [ H(t), \rho ]  + \sum_k \Big[ \hat{c}_k \rho \hat{c}_k\dag - \frac{1}{2} \Big( \hat{c}_k\dag\hat{c}_k \rho + \rho \hat{c}_k\dag \hat{c}_k \Big) \Big], \label{eq:mastereq}
\end{align}
where $\rho$ is the density matrix describing the system at time $t$ and $H(t)$ is the Hamiltonian, the time-dependence of which models the applied gate sequence. The collapse operators $\hat{c}_k$ model incoherent processes. We solve the master equation numerically~\cite{Johansson2013a} in the rotating frame of qubits. 
Incoherent errors are described by Lindblad terms in Eq.~\eqref{eq:mastereq} with
\begin{align}
\hat{c}_{T_1,i} &= \sqrt{\frac{1}{T_{1,i}}} \sigma_{-,i}, \\
\hat{c}_{T_{\phi,i}} &=  \sqrt{ \frac{1}{2} \Big( \frac{1}{T_{2,i}} - \frac{1}{2T_{1,i}} \Big) } \sigma_{z,i},
\end{align}
where $T_{1,i}$ and $T_{2,i}$ are the lifetime and decoherence time (Ramsey decay time) of qubit $i$ as listed in Table~\ref{tab:qb_params}.

\begin{figure}[t] % !b H
\centering
\includegraphics[width=\linewidth]{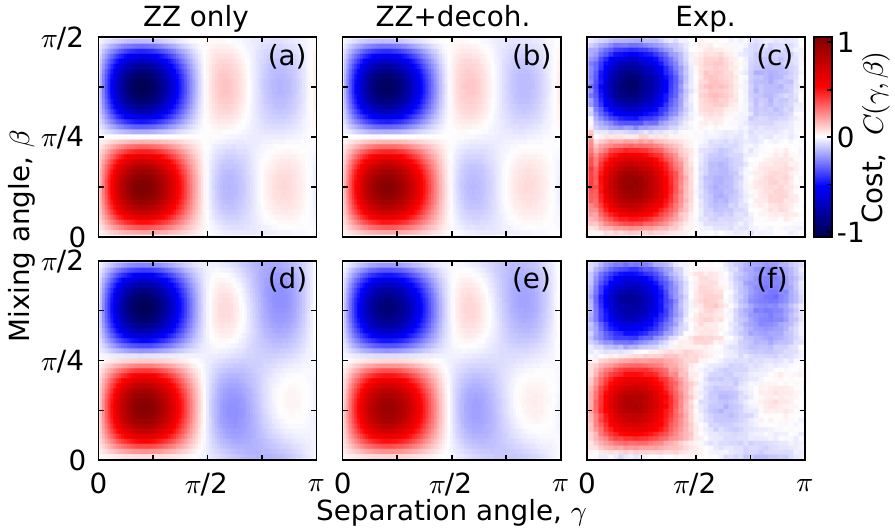}
\caption{Simulated and experimental cost-function landscapes of the three-qubit problem instance for $p=1$. The direct implementation is displayed in (a), (b) and (c). The decomposed version is shown in (d), (e) and (f). The master-equation simulations are performed including  errors from residual $ZZ$-coupling only (a, d), and with both residual $ZZ$-coupling and decoherence (b, e). The experimental data is shown in (c) and (f).}
\label{fig:landscape_sim}
\end{figure}

In addition to the incoherent errors introduced by the Lindblad terms, it is important to also consider the impact of coherent errors  on the algorithm. In our experiment, the main source of coherent errors is residual $ZZ$-coupling between neighboring qubits~\cite{Krinner2020}. To model this coupling in the numerical simulations, we include the Hamiltonian
\begin{equation}
H_{ZZ}/\hbar = \sum_{(i,j)} \alpha_{ij} \ket{11}_{ij}\bra{11}
\end{equation}
where the sum is over connected pairs of qubits with the residual $ZZ$-couplings listed in Appendix~\ref{app:device}. We notice from simulations of the full QAOA circuit that the residual $ZZ$-couplings give rise to the distortions observed in the cost landscapes, see Fig.~\ref{fig:landscape_sim}.
The main effect of decoherence is to reduce the overall contrast of the landscape. In particular, for the direct implementation we find a minimum energy of $-1.04$, $-0.99$ and $-0.98$ for  simulations including only residual $ZZ$-couplings, for  simulations including both residual $ZZ$-couplings and decoherence, and for  experiments, respectively. In comparison, the minimum energy in noise-free simulations is $-1.06$, see Fig.~\ref{fig:landscape}(b).

\section{Seven-qubit problem instance}
\label{app:7qb_landscapes_and_probs}
\begin{figure}[b] % !b H
\centering
\includegraphics[width=\linewidth]{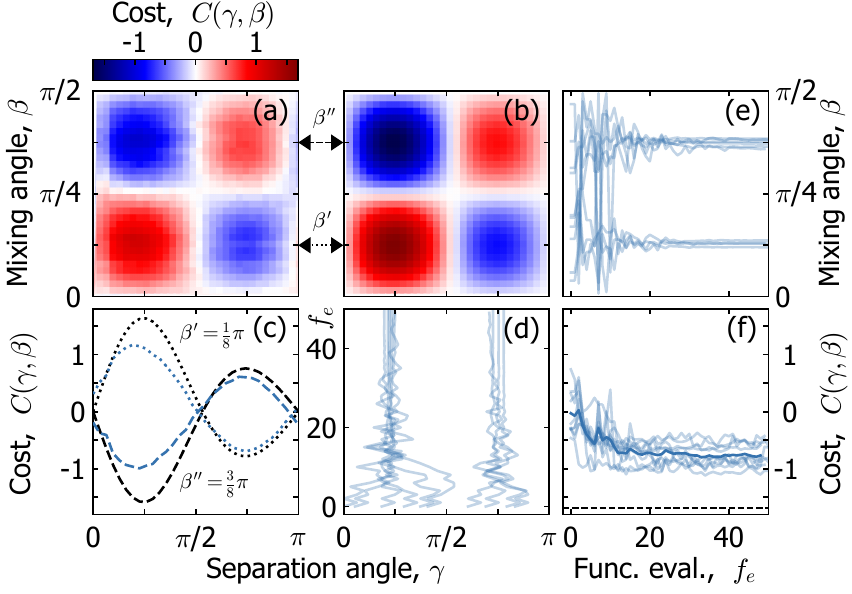}
\caption{Cost function evaluated for $p=1$ on the seven-qubit problem instance, using C-ARB gates. (a) Cost-function landscape as a function of variational parameters measured with direct implementation of C-ARB gates. (b) Cost-function landscape obtained from noise-free simulations. (c) Experimental evaluation (blue) and simulation (black) of the cost function for two horizontal line cuts of (a) and (b), with $\beta'=\pi/8$ (dotted lines) and $\beta''=3\pi/8$ (dashed lines), respectively. (d,e) 10 convergence traces of the separation angle and the mixing angle, respectively, for end-to-end optimization starting from random parameter initialization. (f) Average energy (solid blue line) and individual convergence traces (faded lines) of the energy corresponding to parameters shown in (d,e).}
\label{fig:landscape_7q}
\end{figure}

We use a single-layer QAOA circuit with C-ARB gates to measure the cost-function landscape of the seven-qubit problem instance, see Appendix~\ref{app:pulse} for the full pulse sequence. The measured landscape, see Fig.~\ref{fig:landscape_7q}(a), is in good qualitative agreement with noise-free simulations, see Fig.~\ref{fig:landscape_7q}(b). Due to decoherence, the absolute values of the global extrema are smaller than in noise-free simulations, see Fig.~\ref{fig:landscape_7q}(c).
Starting from random initialization, the convergence traces of the separating angle, the mixing angle and the corresponding cost are displayed in Fig.~\ref{fig:landscape_7q}(d), (e) and (f) respectively.

\begin{figure*}[t] % !b H
\centering
\includegraphics[width=\linewidth]{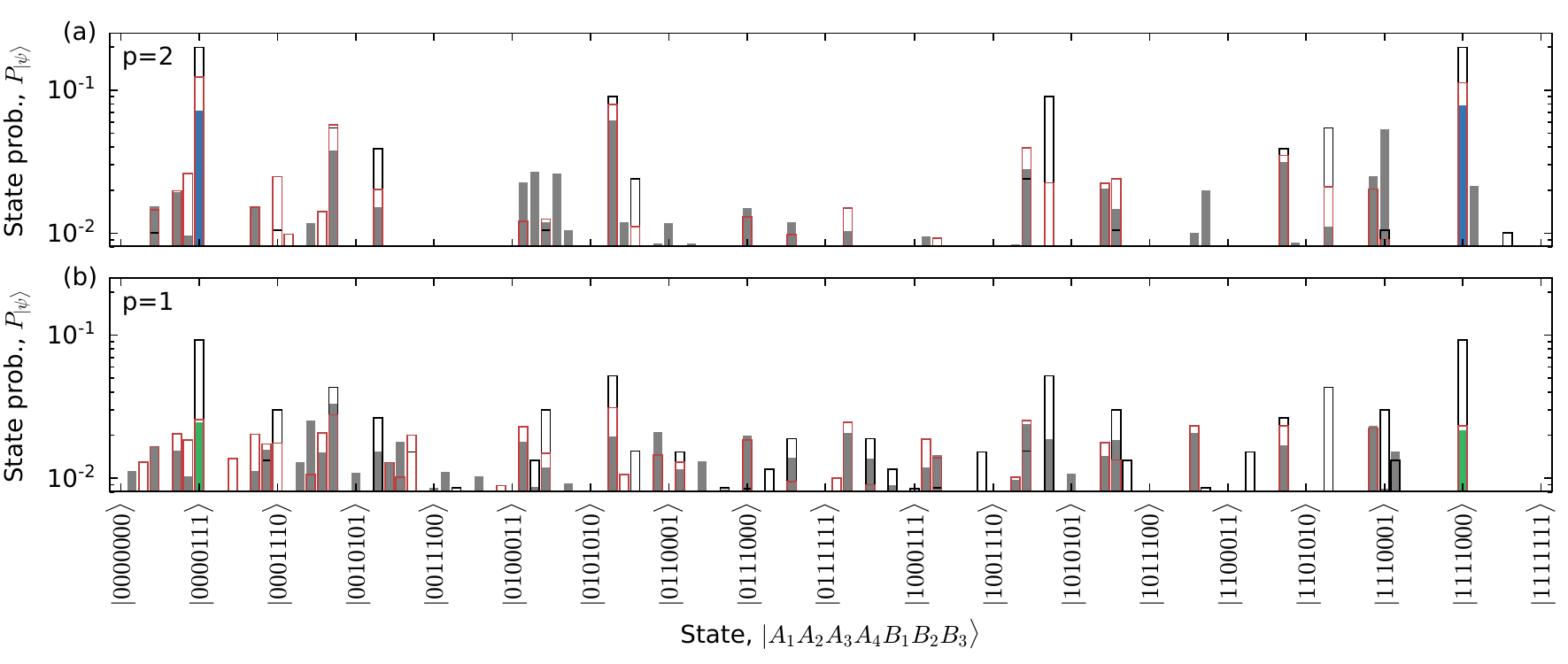}
\caption{Output state probability distribution for the seven-qubit problem instance implemented with C-ARB gates (a) and decomposed using \CZs{} (b). States are measured at optimal parameters for depth of $p=2$ for (a) and  $p=1$  for (b). The filled bars correspond to the measured problem solutions, while the black (red) wire-frames are the expected QAOA outcome from noise-free (master-equation) simulations.}
\label{fig:state_probs_7q}
\end{figure*}

The output state distribution at optimal parameters for the direct and decomposed implementation of C-ARB gates are shown in Fig.~\ref{fig:state_probs_7q}(a) and (b), respectively. We display the distributions yielding highest success probability for each implementation, i.e.\ $p=2$ for the direct implementation and $p=1$ for the decomposed implementation.
For the direct implementation, the two most likely measured states are $\ket{1111000}$ and $\ket{0000111}$, corresponding to the respective selections of subsets $\mathcal{A}$ and $\mathcal{B}$ forming exact covers  of the considered problem instance. 
Conversely, the solution states are not the two most likely measured states for the decomposed implementation. 
For both implementations, the measured data matches well with expectations from master-equation simulations (red wire-frame).

\section{QAOA gate sequences}
\label{app:pulse}
\begin{figure}[t] % !b H
	\centering
	\includegraphics[width=\linewidth]{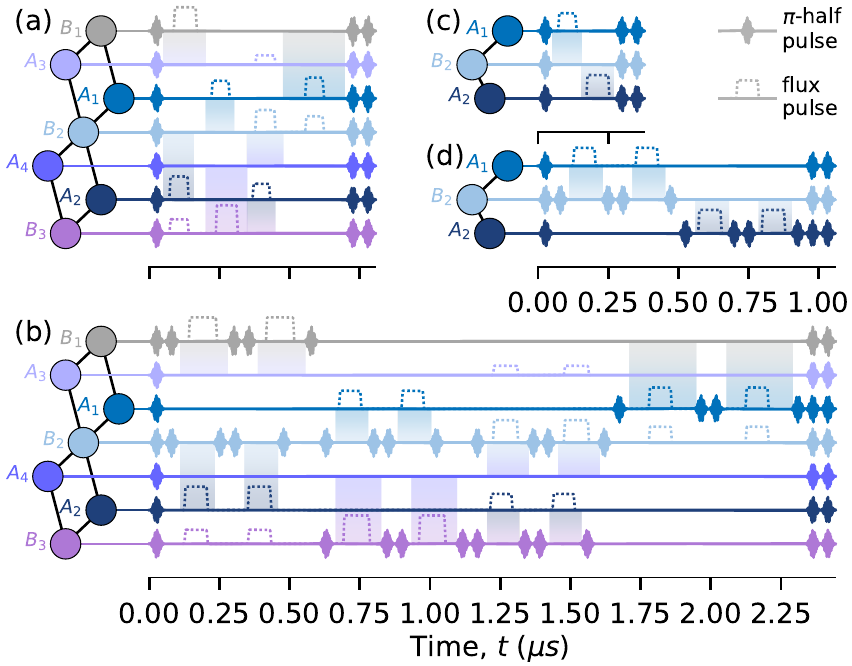}
	\caption{Pulse sequences for implementing QAOA with $p=1$.
		The shaded area around flux pulses illustrates which interaction they implement and how long buffer times before and after the flux pulse are chosen.
		(a) Direct implementation of the seven-qubit instance. (b) Decomposed implementation of the seven-qubit instance.
		(c) Direct implementation of the three-qubit instance. (d) Decomposed implementation of the three-qubit instance.
	}
	\label{fig:pulse_seq}
\end{figure}

Fig.~\ref{fig:pulse_seq}(a) shows the pulse sequence generated by the AWGs for a single layer of QAOA in the seven-qubit instance using the direct implementation of C-ARB gates.
Since the length of the flux pulses depends on the required phase, see Fig.~\ref{fig1}(c) and (d),
the pulse sequence is shown for a representative flux pulse length (close to the average) that we obtain for $\gamma=\frac{\pi}{5}$.
After an initial $\pi/2$-pulse on each qubit to prepare a $\ket{+}^{\otimes 7}$ state,
the phase-separation operator $U_{C}$ of the first QAOA layer starts with two parallel C-ARB gates corresponding to the couplings $J_{A_3B_1}$ and $J_{A_2B_2}$,
while qubit $B_3$ is detuned by an additional flux pulse to avoid an unwanted interaction when the ef-transition frequency of $A_2$ crosses the parking frequency of qubit $B_3$.
Since the additional $Z_{\Gamma_{ij}}$ rotations, see Fig.~\ref{fig1}(a), are implemented as virtual gates~\cite{McKay2016a} through a redefinition of the reference frame,
they are not shown in the pulse sequence.
After the last round of flux pulses,
the final two $\pi/2$-pulses for each qubit (plus a virtual gate between them) implement the mixing operator $U_B=\e{-\i \beta B}$, 
where we have decomposed each term $\e{-\i \beta X_i}$ as shown in Fig.~\ref{fig1}(a).
After the end of the shown pulse sequence, we perform qubit readout.

In the significantly longer pulse sequence shown in Fig.~\ref{fig:pulse_seq}(b), the controlled arbitrary phase gates are decomposed as described in Fig.~\ref{fig1}(b).
Each Hadamard gate is implemented by a $\pi/2$-pulse and a $Z_\pi$ rotation via a virtual gate, and the $Z_{\Gamma_{ij}}$ in the center of the gate decomposition is another virtual gate.
Pulse sequences for the direct and the decomposed implementation of the three-qubit problem instance are shown in Fig.~\ref{fig:pulse_seq}(c) and (d), where analogous explanations apply.

To implement additional layers, the pulses between the end of the initialization pulses and the start of the readout are repeated $p-1$ times.
For the configurations considered in Fig.~\ref{fig:success}, this leads to the gate counts shown in Table~\ref{tab:gate_count}.

\begin{table}[b]
\centering
\caption{Number of two-qubit gates (first row), single-qubit gates (second row), and virtual gates (third row) in the implemented QAOA sequences.}
{\def\arraystretch{1.25}%
	\begin{tabular}{ l c c c c c c c c c }\hline
		\# of layers, $p$ & 1 & 2 & 3 & 4 & 5 & 6 & 7 & 8 & 9 \\\hline\hline
		3 qb direct & \parbox{5mm}{\vspace{1mm}2\\9\\7\vspace{1mm}} & \parbox{5mm}{\vspace{1mm}4\\15\\14\vspace{1mm}} & \parbox{5mm}{\vspace{1mm}6\\21\\21\vspace{1mm}} & \parbox{5mm}{\vspace{1mm}8\\27\\28\vspace{1mm}} & \parbox{5mm}{\vspace{1mm}10\\33\\35\vspace{1mm}} & \parbox{5mm}{\vspace{1mm}12\\39\\42\vspace{1mm}} & \parbox{5mm}{\vspace{1mm}14\\45\\49\vspace{1mm}} & \parbox{5mm}{\vspace{1mm}16\\51\\56\vspace{1mm}} & \parbox{5mm}{\vspace{1mm}18\\57\\63\vspace{1mm}} \\\hline
		3 qb decomposed & \parbox{5mm}{\vspace{1mm}4\\17\\13\vspace{1mm}} & \parbox{5mm}{\vspace{1mm}8\\31\\26\vspace{1mm}} & \parbox{5mm}{\vspace{1mm}12\\45\\39\vspace{1mm}} & \parbox{5mm}{\vspace{1mm}16\\59\\52\vspace{1mm}} & \parbox{5mm}{\vspace{1mm}20\\73\\65\vspace{1mm}} & \parbox{5mm}{\vspace{1mm}24\\87\\78\vspace{1mm}} & \parbox{5mm}{\vspace{1mm}28\\101\\91\vspace{1mm}} & \parbox{5mm}{\vspace{1mm}32\\115\\104\vspace{1mm}} & \parbox{5mm}{\vspace{1mm}36\\129\\117\vspace{1mm}} \\\hline
		7 qb direct & \parbox{5mm}{\vspace{1mm}7\\21\\21\vspace{1mm}} & \parbox{5mm}{\vspace{1mm}14\\35\\42\vspace{1mm}} & \parbox{5mm}{\vspace{1mm}21\\49\\63\vspace{1mm}} & \parbox{5mm}{\vspace{1mm}28\\63\\84\vspace{1mm}} & \parbox{5mm}{\vspace{1mm}35\\77\\105\vspace{1mm}} & \parbox{5mm}{\vspace{1mm}42\\91\\126\vspace{1mm}}&&& \\\hline
		7 qb decomposed & \parbox{5mm}{\vspace{1mm}14\\49\\42\vspace{1mm}} & \parbox{5mm}{\vspace{1mm}28\\91\\84\vspace{1mm}} & \parbox{5mm}{\vspace{1mm}42\\133\\126\vspace{1mm}} & \parbox{5mm}{\vspace{1mm}56\\175\\168\vspace{1mm}}&&&&& \\\hline
\end{tabular}}
\label{tab:gate_count}
\end{table}

\section{Post-selection} \label{app:post_selection}
For both QAOA implementations, we discard all measured states containing at least one leakage event. We show the percentage of single-shot measurements we keep as a function of the number of layers in the top half of Table~\ref{tab:post_selection}. We estimate the corresponding average leakage per gate as $\lambda \approx 1 - P_{\textrm{post}}^{1/n_\textrm{g}}$, where $P_{\textrm{post}}$ is the fraction of data left after post-selection and $n_\textrm{g}$ is the number of two-qubit gates in the sequence. All average leakage per gate values lie between 0.2\% and 2.1\%.

\begin{table}[t]
\centering
\caption{(First 4 rows) Percentage of data kept after discarding all measured states containing at least one leakage event. (Bottom 4 rows) Corresponding average leakage per two-qubit gate in percent.}
\begin{tabular}{lccccccccc}
\hline
\# of layers, $p$ & 1 & 2 & 3 & 4 & 5 & 6 & 7 & 8 & 9\\
 \hline
 \hline
3 qb direct & 99.2 & 97.1 & 97.9 & 97.1 & 96.4 & 97.1 & 96.6 & 94.1 & 93.6 \\
3 qb dec. & 98.6 & 98.3 & 96.4 & 92.1 & 95.8 & 94.0 & 94.8  & 94.4 & 93.5 \\
7 qb direct & 87.1 & 87.7 & 80.6 & 55.9 & 79.2 & 77.5 & & & \\
7 qb dec. & 80.2 & 82.4 &  61.6 &  62.1 & & & & & \\
\hline
3 qb direct & 0.4 & 0.7&  0.3& 0.4& 0.4 &0.2& 0.2 &0.4& 0.4\\ 
3 qb dec. & 0.4 &0.2& 0.3& 0.5& 0.2& 0.3& 0.2& 0.2& 0.2\\ 
7 qb direct & 1.9 & 0.9 & 1.0 & 2.1 & 0.7 & 0.6 & & & \\
7 qb dec. & 1.6 & 0.7 & 1.1 & 0.8 & & & & & \\
\hline
\end{tabular}
\label{tab:post_selection}
\end{table}

\end{appendix}

%----------------------------------------------------------------------------------------
%	BIBLIOGRAPHY
%----------------------------------------------------------------------------------------

\bibliography{QudevRefDB, references}

\end{document}